\documentclass[twocolumn,showpacs,preprintnumbers,amsmath,amssymb,aps,prb,10pt]{revtex4-1}

\usepackage{graphicx}
\usepackage{subfigure}
\usepackage{bm}

\newcommand{\bi}{\begin{itemize}}
\newcommand{\ei}{\end{itemize}}
\newcommand{\be}{\begin{eqnarray}}
\newcommand{\ee}{\end{eqnarray}}
\newcommand{\nn}{\nonumber}
\newcommand{\m}{\mathbf}

\begin{document}

\title{Effects of bond disorder and surface amorphization on optical phonon lifetimes and Raman peak shape in crystalline nanoparticles}

\author{Oleg I. Utesov$^{1,2,3}$}
\email{utiosov@gmail.com}

\author{Sergei V. Koniakhin$^{4,5}$}
\email{kon@mail.ioffe.ru}

\author{Andrey G. Yashenkin$^{1,2}$}

\affiliation{$^1$Petersburg Nuclear Physics Institute NRC ``Kurchatov Institute'', Gatchina 188300, Russia}
\affiliation{$^2$Department of Physics, St. Petersburg State University, St.Petersburg 199034, Russia}
\affiliation{$^3$St. Petersburg School of Physics, Mathematics, and Computer Science, HSE University, 190008 St. Petersburg, Russia}
\affiliation{$^4$Institute Pascal, PHOTON-N2, University Clermont Auvergne, CNRS, 4 Avenue Blaise Pascal, Aubi\`{e}re Cedex 63178, France}
\affiliation{$^5$St. Petersburg Academic University - Nanotechnology Research and Education Centre of the Russian Academy of Sciences, St.Petersburg 194021, Russia}

\date{\today}

\begin{abstract}


Optical phonons  in nanoparticles with the randomness of interatomic bonds are considered both analytically and numerically. For weak dilute disorder two qualitatively different regimes of separated and overlapped levels are observed, resembling the case of random atomic masses investigated previously. At stronger and/or more dense disorder, the particles become essentially inhomogeneous, thus constituting the minimal model to describe an amorphous phase, where the picture of vibrational modes becomes more subtle. We concentrate here on the experimentally relevant case of   strong disorder located  near the particle surface and formulate the core-shell model aimed to describe the ubiquitous phenomenon of particle surface amorphization. We observe a peculiar effect of volume optical phonons ``repelling'' from the disordered shell. It results in the Raman spectrum in the form of a combination of narrow well-resolved peaks stemming from the  quantized modes of a pure particle core (red-shifted due to its effective smaller size), and the noisy background signal from the disordered shell placed primarily to the right from the main Raman peak.

\end{abstract}

\maketitle

\section{Introduction}
\label{SIntro}

Nanoparticles find numerous applications in modern science and technology \cite{bhatia2016nanoparticles,khan2019nanoparticles}, which span the areas from composite materials~\cite{yadav2020polysulfone,qaid2020optical,jiang2020corrosion,yu2020fabrication} and  state-of-the-art thermal management~\cite{yoshida2004thermal,kidalov2009thermal,sun2021influence} to quantum computing~\cite{riedel2017deterministic,andrich2017long} and biological sciences~\cite{faklaris2009photoluminescent,rahban2010nanotoxicity,kurdyukov2019fabrication,popescu2020efficient,rehman2020nanodiamond,berdichevskiy2021comprehensive}. Nanoparticles can be classified by the type of constituting material as metallic/semiconducting/dielectric/polymer, etc. Semiconducting and dielectric species could exist in crystalline and amorphous forms, depending on the arrangement of atoms. An intermediate and very widespread case is the nanoparticles with crystalline core and amorphous shell (or coating). In this context, it is instructive to mention the nanodiamonds with sp$^3$ hybridized core carbon atoms and sp$^{3-x}$ bonded carbon atoms within the amorphous or partly graphitized shell~\cite{aleksenskii1997diamond,tomita2002diamond,tomita2002optical,mykhaylyk2005transformation,zou2010transformation,dideikin2017rehybridization,PhysRevResearch.2.013316,kryshtal2021primary}. Importantly, the fraction of surface atoms in nanoparticles is comparable with the one of the core. E.g., the 4 nm diamond particle contains approx. 6000 atoms, 3000 of which are on the surface (within the $a_0$ thickness layer, where $a_0$ is the diamond lattice parameter). Therefore, \textit{the nanoparticle properties are significantly influenced by its surface}.

This paper continues our efforts to build up the detailed and precise  microscopic theory of  Raman scattering in nanoparticles \cite{ourDMM,ourEKFG,our3,our4,ourshort}.  It is focused on two main goals. First, it
elucidates the role of strongly disordered surface shell (surface amorphization) for the Raman response in crystalline nanoparticles.
Our main result here is the demonstration of possibility to identify and to separate the Raman contribution originating  from the disordered surface shell and the one which comes from the relatively clean core of the particle. We call it the ``core-shell model'', despite usually this term is used for particles with a coating of another material (see, e.g., Ref.~\cite{dzhagan2009influence}).

The second goal is to extend the microscopic theory of the optical phonon scattering in weakly disordered nanoparticles previously developed for disorder in the form of random atomic masses \cite{our3,our4} onto the randomness of interatomic bonds. Both these mechanisms are definitely mutually related (an atom with different mass embedded into the lattice is typically connected to its neighbours by the bonds with different strength) and coexist in real nanocrystallites. Here we report the total similarity of contributions stemming  these two mechanisms as far as the properties of optical phonons in nanoparticles in harmonic approximation are concerned.

In order to proceed with the above program, we utilize the DMM-BPM \cite{ourDMM} and the EKFG \cite{ourEKFG} methods previously developed for the description of Raman spectra (RS) of {\it clean} crystalline nanoparticles. These methods allow to describe quantitatively the principal features of nanoparticle RS, including the shift of the main Raman peak to lower frequencies as a consequence of the size quantization effect~\cite{meilakhs2016new} and the asymmetry of the peak shape, appearing due to ``Raman-active'' modes with frequencies lower than the highest optical phonon mode frequency. Both DMM-BPM and EKFG approaches were shown to reproduce the RS of {\it nanocrystals}   better, than the famous Phonon Confinement Model (PCM)~\cite{richter1981one,campbell1986effects} and without introducing new fitting parameters \cite{osswald2009phonon,korepanov2017,korepanov2017quantum,korepanov2020localized,stehlik2015size}. Later on, DMM-BPM method has been successfully used in Ref.~\cite{ranipredicting} for the description and analysis of {\it amorphous} Si Raman data.

We incorporate disorder into the treatment, properly adopting the microscopic theory of Refs.~\cite{our3,our4} for the description of bond disorder. Similarly to the random masses problem investigated in Refs.~\cite{our3,our4}, this case allows an analytical treatment for dilute weak (Born) and strong disorder, whereas the dense impurities as well as the surface randomness are a subject of numerical studies.

\begin{figure}[tbp]
\includegraphics[width=0.45\linewidth]{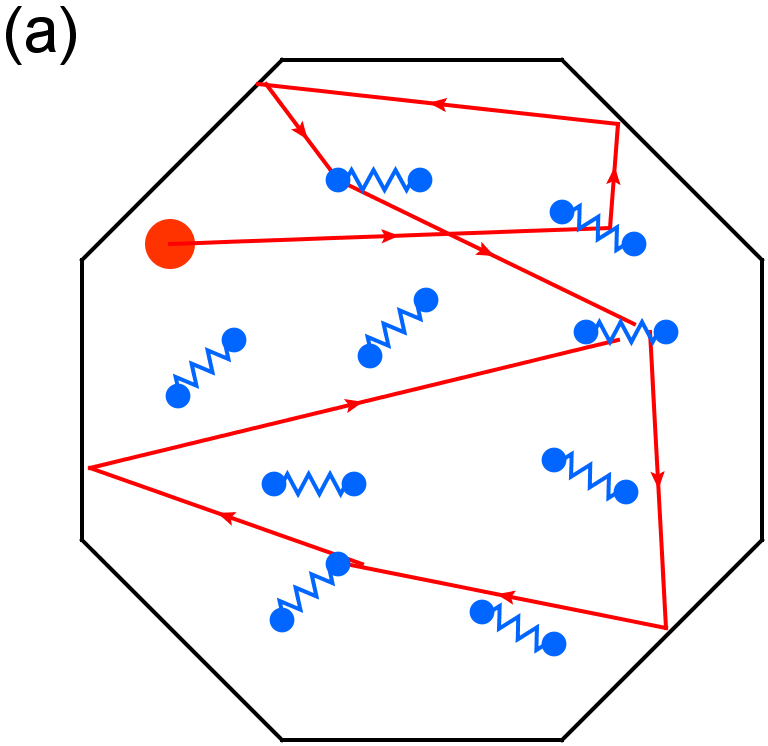} \includegraphics[width=0.45\linewidth]{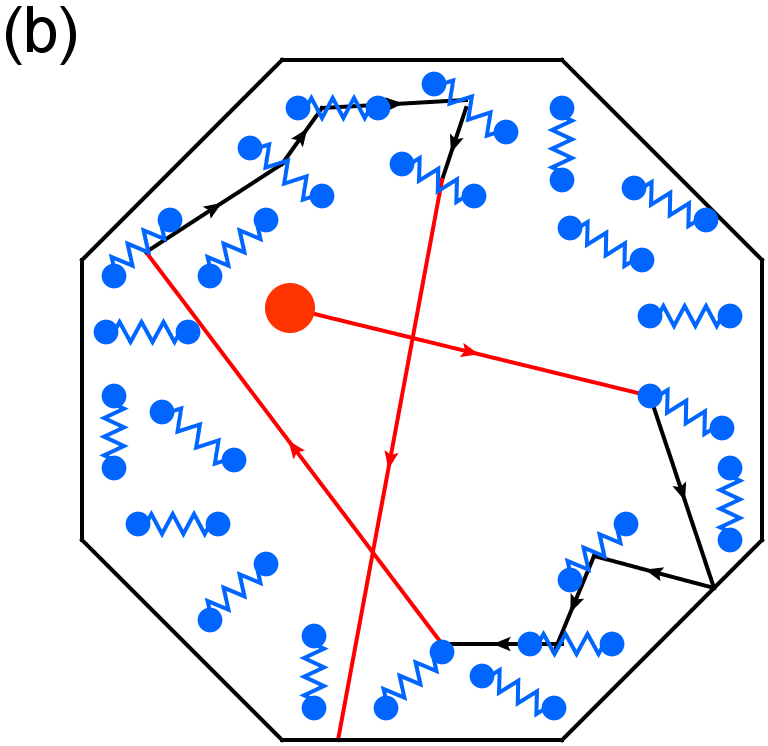}
\caption{A sketch illustrating the problem addressed in the present paper. Bond disorder leads to optical phonons scattering off the defects, which results in finite phonon linewidths. (a) Model, where defect bonds distributed over the \textit{entire} nanoparticle volume. (b) Proposed model for theoretical description of nanoparticle surface amorphization with strongly disordered shell, whereas the core is pure. As it is shown in the present paper, core ``volume'' modes cannot propagate in the shell and are pushed away from its region. }
\label{Sketch}
\end{figure}


Investigating the model of Gaussian uncorrelated disorder at small values of the disorder strength parameter $S$, we find two regimes of the optical phonon line broadening, namely, the regime of separated levels with the broadening which has the form $\Gamma \sim \sqrt{S}/L^{3/2}$, and the overlapped levels regime characterized by larger the linewidths $\Gamma \sim S /L$, where $L$ is the particle size. The difference with the result of Refs.~\cite{our3,our4} is in the meaning of the parameter $S$, which is now the product of the defective bonds concentration and the (squared normalized) mean bond defect amplitude (see Eq.~\eqref{dis1} below).  At larger $S$ the localized phonon-impurity bound states on disordered bonds emerge, and the resonant scattering phenomenon comes into play. As a result, the line broadening is significantly enhanced, and its $L$-dependence is altered. At yet larger $S$ we focus on the surface amorphization. We observe
 the peculiar behaviour of volume modes: they can be considered as weakly broadened quantized modes of a pure nanoparticle core. The Raman spectrum in this regime can be represented as a superposition of well-resolved narrow peaks stemming from the quantized modes of the particle core and the noisy background originating from the amorphous shell.


It means that the interpretation of  Raman data for such coated particles with the use of the approach developed in Ref.~\cite{ourshort} or by means of the powder X-Ray diffraction (XRD)~\cite{ozerin2008x,abdullahi2020explosive,rabiei2020comparing} will reveal the parameters (mean size, size distribution function variance, disorder strength, shape, etc.) of particle cores rather than that of coated particles, so these approaches should be properly modified to account for the coating.
On the contrary, the dynamic light scattering (DLS) and the atomic force microscopy (AFM) provide us with the information about the total nanoparticle size including the surface shell~\cite{PhysRevResearch.2.013316,kovavrik2020particle,stehlik2021size}. The high resolution transmission electron microscopy (HRTEM) is suitable for the insight into atomic structure of particles, albeit it is a strictly local technique~\cite{stehlik2016high,dideikin2017rehybridization,kryshtal2021primary,stehlik2021size}.

The rest of the paper is organized as follows. Section~\ref{STheor} contains the description of our model and the analysis of the applied theoretical concept. In Sec.~\ref{SRes} we present our results for both the bond disorder distributed over the entire particle volume and for the core-shell model, and discuss the optical phonon linewidths and the corresponding Raman spectra. We formulate our conclusions in Sec.~\ref{SCon}. Some important details of EKFG calculations are presented in Appendix.

\section{Theory}
\label{STheor}

In this section we briefly discuss theoretical concepts important for our treatment and adapt them for bond disorder. The detailed presentation of this theory can be found in Ref.~\cite{our3}

Within the harmonic approximation vibrational subsystem of a nanoparticle can be described with the use of the ``balls-springs'' model
\begin{equation}\label{hgen0}
  {\cal H} = \sum_l \frac{p^2_l}{2 m_l} +  \frac{1}{2} \, \sum_{\langle l l^{\prime}\rangle} K_{ll^{\prime}} \left( \mathbf{r}_l - \mathbf{r}_{l^{\prime}} \right)^2.
\end{equation}
Here the first summation includes all sites and the second one runs over all pairs of (neighboring) atoms.

Disorder can be incorporated into this Hamiltonian in several ways. In the present research we focus on the bond disorder effects (random atomic masses have been studied in Refs.~\cite{our3,our4}). We consider (bearing in mind, e.g., the nanodiamonds) the particles each containing $N$ atoms of equal mass $m$ (generalization to the system with different site masses is straightforward). Bond rigidities can deviate from their value in the pure system. Without any additional knowledge about chemical properties of the nanoparticle and its surface, it is natural to assume the delta-correlated Gaussian character of the bond disorder:
\be \label{dis1}
  \frac{\langle \delta K_{l_1 l_1^{\prime}} \delta K_{l_2 l_2^{\prime}} \rangle}{\langle K_{l_1 l_1^{\prime}}\rangle^2} = S \delta_{l_1 l_1^{\prime},l_2 l_2^{\prime}}
\ee
Such an assumption (white-noise disorder) is widely used for the description of disordered condensed matter systems universal properties, without specifying particular information about the disorder.

For disordered particle shell we modify the definition~\eqref{dis1} as follows:
\be \label{dis2}
  \frac{\langle K_{l_1 l_1^{\prime}} K_{l_2 l_2^{\prime}} \rangle_{\text{shell}}}{ \langle K_{l_1 l_1^{\prime}}\rangle^2} = S_{\rm shell} \delta_{l_1 l_1^{\prime},l_2 l_2^{\prime}},
\ee
where the averaging takes place over the disordered shell area of the particle. This definition reflects the real strength of disorder in the shell, not smearing it over the entire volume of the particle (cf. Eq.~\eqref{dis1}). Moreover, we shall model an amorphous (strongly disordered) surface shell using this model at large $S_{\rm shell}$. Previously, similar model has been used for theoretical study of low-energy (acoustic and diffusive) modes in bulk amorphous solids~\cite{beltukov2013,conyuh2021}.

\subsection{EKFG}

The easiest way to demonstrate the equivalence between the present case (bond disorder) and the model with disorder in atomic masses is to use the continuous EKFG approach~\cite{ourEKFG}. Within the EKFG framework, the long-wavelength optical phonon eigenfunction $Y$ satisfies the following equation:
\be \label{EKFG1}
(\partial^2_t + C_1 \Delta + C_2) Y =0
\ee
with the Dirichlet boundary conditions $Y|_{\partial \Omega} = 0$. So, the optical phonon spectrum has the form:
\be \label{EKFG2}
  \omega^2_q = C_2 - C_1 q^2,
\ee
where the ``momentum'' $q$ can be found from the boundary problem
\be \label{EKFG3}
 \Delta Y + q^2 Y = 0, \quad Y|_{\partial \Omega} = 0.
\ee
Importantly, the model parameters $C_{1,2}$ are interconnected with the parameters of the long-wavelength optical phonon spectrum, which can be approximated as
\be
  \omega_\m{q} = \omega_0 - \alpha q^2,
\ee
and can be obtained, e.g., from the Keating model~\cite{keating1966effect}.

The eigenfunctions and eigenfrequencies of the above problem can be used for the optical phonon field quantization as follows:
\begin{equation}\label{Phi1}
  \Phi(\mathbf{r}) = \sum_n \frac{1}{\sqrt{2 \omega_n}} \left( b_n Y_n(\mathbf{r}) + b^\dagger_n Y^*_n(\mathbf{r}) \right),
\end{equation}
and the corresponding momentum operator reads
\begin{equation}\label{Pi1}
  \Pi(\mathbf{r})=  - \sum_n i \sqrt{\frac{\omega_n}{2}} \left( b_n Y_n(\mathbf{r}) - b^\dagger_n Y^*_n(\mathbf{r}) \right).
\end{equation}
Using the last two equations, one can introduce the Hamiltonian of the EKFG model
\begin{equation}\label{Ham1}
  \mathcal{H}_0 = \frac{1}{2} \int_\Omega d^3 \mathbf{r} \left[ \Pi^2 - C_1 (\nabla \Phi)^2 + C_2 \Phi^2 \right].
\end{equation}
Its quantized version is simply
\begin{equation}\label{Ham2}
  \mathcal{H}_0 = \sum_n \omega_n (b^\dagger_n b_n +1/2).
\end{equation}

As in the case of random atomic masses, the disorder-induced perturbation of the Hamiltonian~\eqref{Ham2} reads
\begin{equation}\label{Pert1}
  \mathcal{H}_{imp} = \frac{1}{2} \int_\Omega d^3 \mathbf{r} \, \left( C_2(\mathbf{r}) - \overline{C_2} \right) \, \Phi^2,
\end{equation}
but now with $C_2(\mathbf{r}) - \overline{C_2} = \omega^2_0 \delta K(\mathbf{r})/K $ ($\overline{C_2} = \omega^2_0$). Comparing this formula with the previous result of Refs.~\cite{our3,our4} ($C_2(\mathbf{r}) - \overline{C_2} = - \omega^2_0 \delta m(\mathbf{r})/m $), we conclude that these two problems has one-to-one correspondence (within the EKFG approach validity domain $q a_0 \ll 1$). This results in the phonon lines broadening of the form
\be
  \label{eq_separated}
  \Gamma_n \propto \sqrt{\frac{S}{L^3}}
\ee
in the separated levels regime, and
\be
  \label{eq_overlapped}
  \Gamma_n \propto S q_n
\ee
or equivalently $\Gamma_n \propto \nu_n(p) S / L$ if the optical phonon spectral weights are essentially overlapped. Here $\nu_n(p)$ is the coefficient which depends on the particle shape and strongly depends on the phonon mode quantum number $n$. However, the growth of $\Gamma_n$ with $n$ in the bond disorder model is significantly suppressed in small nanoparticles as compared to random atomic masses, where  $\Gamma_n \approx n \Gamma_1$. Importantly, the EKFG approach becomes only qualitatively correct when the phonon momentum is large enough. In order to discuss this suppression, we turn to the description in terms of the discrete balls-springs (DMM) model.

\subsection{Dynamical matrix and Green's functions}

Within the framework of the dynamical matrix method (DMM) we shall perform the quantization in the basis of pure particle eigenfunctions $\m{Y}_n(\m{R}_l)$ (n=1...3N). Then the displacement and momenta operators obtain the following form:
\begin{equation}\label{rlQ}
  \mathbf{r}_l = \frac{1}{\sqrt{2 m}} \sum_n \frac{\mathbf{Y}_n(\mathbf{R}_l)}{\sqrt{\omega_n}} (b_n + b^\dagger_n)
\end{equation}
and
\begin{equation}\label{plQ}
  \mathbf{p}_l = \frac{i \sqrt{m}}{\sqrt{2}} \sum_n \mathbf{Y}_n(\mathbf{R}_l) \sqrt{\omega_n} (b^\dagger_n - b_n),
\end{equation}
correspondingly.

Next, we can split the quantized Hamiltonian~\eqref{hgen0} onto the bare part
\be
  \mathcal{H}_0 = \sum_n \omega_n (b^\dagger_n b_n + 1/2)
\ee
and the perturbation
\be \label{himp}
  \mathcal{H}_{imp} &=& \frac{1}{4m} \sum_{n,n^\prime, \langle l l^\prime \rangle} \frac{\delta K_{l l^\prime}}{\sqrt{\omega_n \omega_{n^\prime}}} (b_n +b^\dagger_n)(b_{n^\prime}+b^\dagger_{n^\prime}) \\ && (\mathbf{Y}_n(\mathbf{R}_l)-\mathbf{Y}_n(\mathbf{R}_{l^\prime})) \cdot (\mathbf{Y}_{n^\prime}(\mathbf{R}_l)-\mathbf{Y}_{n^\prime}(\mathbf{R}_{l^\prime})). \nn
\ee
Then, we can formulate the conventional disordered diagram technique for operators $\phi_n = b_n +b^\dagger_n $ and the Green's functions $-i \langle \hat{T} \phi_n \phi_n \rangle$, where $\hat{T}$ is the time-ordering operator. The Green's function averaged over disorder configurations reads
\begin{equation}\label{DQ}
  D_n(\omega)= \frac{2  \omega_n}{\omega^2 - \omega^2_n - 2 \omega_n \Pi_n (\omega)},
\end{equation}
where we introduce the self-energy part $\Pi_n (\omega)$, which can be calculated in different approximations (see Ref.~\cite{our3}). For instance, at $S \ll 1$ (the self-consistent) Born approximation depicted in Fig~\ref{Diagrams}(a) can be used (see  Appendix for some extra details).

\begin{figure}[tbp]
\includegraphics[width=0.99\linewidth]{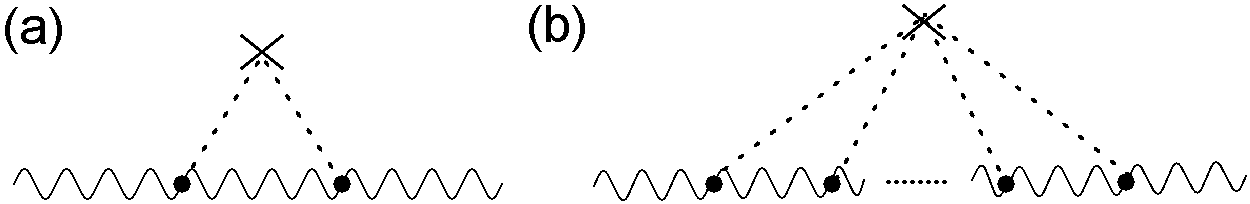}
\caption{ The diagrams relevant for the discussion of optical phonon lines broadening in weakly disordered particles are the following . (a) The Born approximation for a weak Gaussian disorder. In the case of separated levels regime the Green's function under the ``arch'' is dressed by disorder (self-consistent Born approximation). (b) For rare strong impurities the T-matrix approach is fruitful, which, e.g., allows to discuss the phonon states localized on the impurities. If the impurities are strong and their concentration is significant (for instance, in the amorphous shell model) both approaches fail. }
\label{Diagrams}
\end{figure}

At the first glance, the perturbation~\eqref{himp} is essentially different from the one stemming from the disorder in atomic masses. It is indeed the case for acoustic phonons, where
\be
|\mathbf{Y}_n(\mathbf{R}_l)-\mathbf{Y}_n(\mathbf{R}_{l^\prime})| \sim |\mathbf{Y}_n(\mathbf{R}_l)| q_n a_0 \ll |\mathbf{Y}_n(\mathbf{R}_l)| \quad
\ee
in the long-wavelength limit. However, for long-wavelength optical phonons and neighboring atoms $\mathbf{Y}_n(\mathbf{R}_l) \approx -\mathbf{Y}_n(\mathbf{R}_{l^\prime})$ and Eq.~\eqref{himp} can be approximately rewritten in the form:
\be \label{himp2}
  \mathcal{H}_{imp} &=& \frac{1}{m} \sum_{n,n^\prime, \langle l l^\prime \rangle} \frac{\delta K_{l l^\prime}}{\sqrt{\omega_n \omega_{n^\prime}}} (b_n +b^\dagger_n)(b_{n^\prime}+b^\dagger_{n^\prime}) \\ && \mathbf{Y}_n(\mathbf{R}_l) \cdot \mathbf{Y}_{n^\prime}(\mathbf{R}_l). \nn
\ee
Corrections to this expression are small in the parameter $q_n a_0$ . Nevertheless, they  become important even for phonon lines with numbers $n \sim 1$ in very small nanoparticles, leading to diminishing of the broadening for these lines. As a reasonable estimation one can use for short wavelength optical phonons $|\mathbf{Y}_n(\mathbf{R}_l)-\mathbf{Y}_n(\mathbf{R}_{l^\prime})| \approx |\mathbf{Y}_n(\mathbf{R}_l)| $ .

We see that Eq.~\eqref{himp2} once again demonstrates the similarity between the case of disorder in atomic masses and the one of disorder in bond rigidities for optical phonons.

\begin{figure}[tbp]
\includegraphics[width=0.99\linewidth]{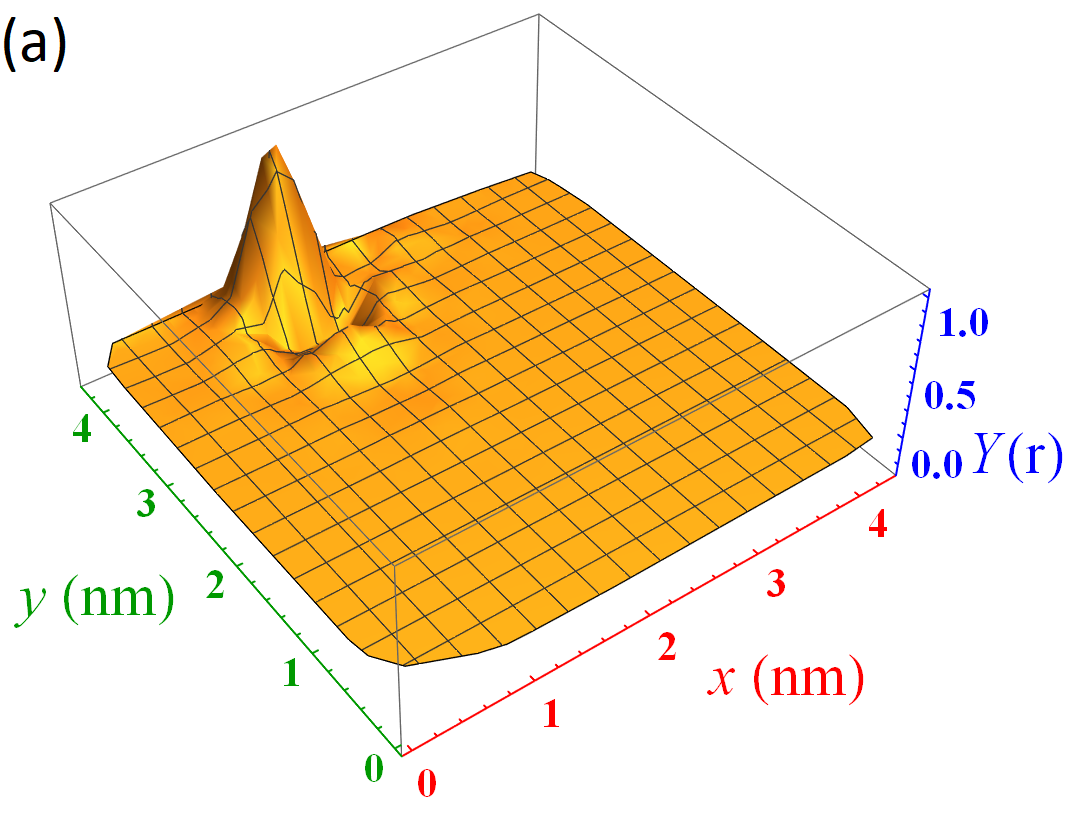}
\includegraphics[width=0.99\linewidth]{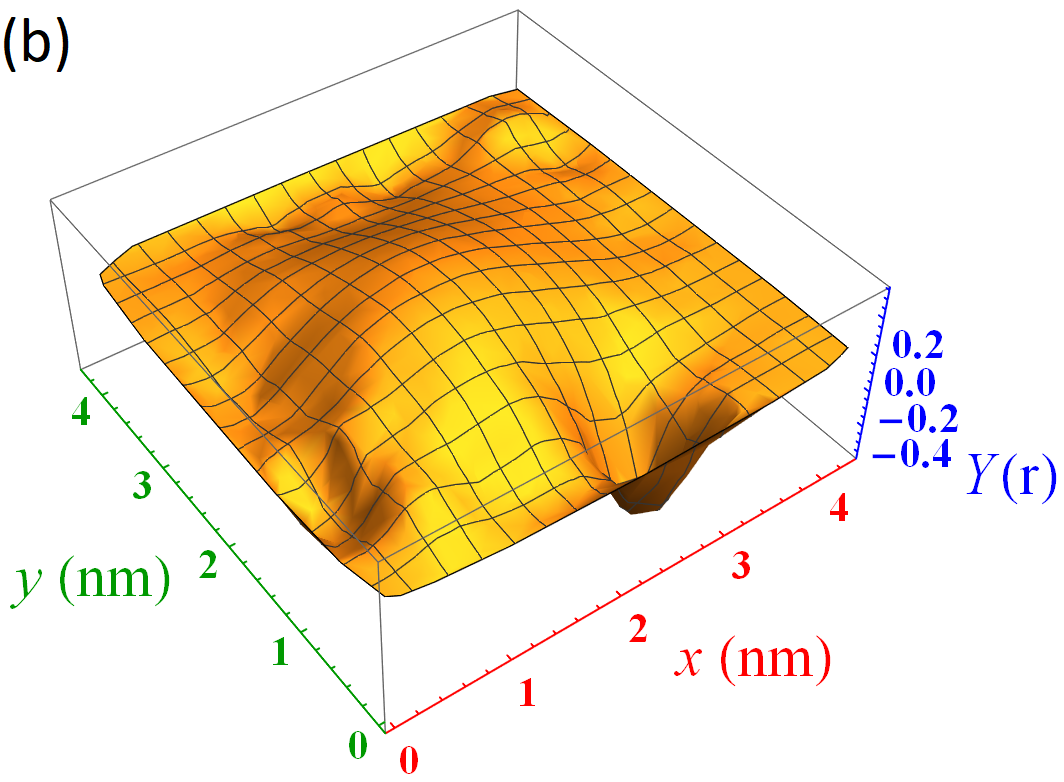}
\caption{EKFG eigenfunctions of a cubic 5.3 nm diamond particle with strongly disordered $S =0.09$ ``amorphous'' shell. The core-to-size ratio is $L_0/L = 5/6$. (a) The localized state in the shell has a sharp peak shape. (b) The eigenfunction ``repelled'' from the disordered shell, which is very similar to the highest optical phonon mode of the pure core. Its frequency corresponds to the core size $L_0$, and in average its broadening is due to the rare regions in the shell, where the eigenfunction amplitude is not negligible.  }
\label{Waves}
\end{figure}

\subsection{Strong impurities. Resonant scattering and localized states}

At large $S$ there is a significant amount of strong impurities, which can host the localized optical vibration modes with $\omega > \omega_0$. In the one impurity problem this possibility is governed by the dimensionless impurity strength $U$; the corresponding analysis can be performed using the conventional T-matrix approach (see Fig.~\ref{Diagrams}(b)). For disordered atomic masses it reads (see Ref.~\cite{our3})
\be \label{Um}
  U = \frac{\delta m}{m + \delta m}.
\ee
while in the present (random bonds) model 
\be \label{Uk}
  U = \frac{\delta K}{K},
\ee
Another difference arises since Eq.~\eqref{Uk} enters the theory with the opposite sign as compared to Eq.~\eqref{Um}. So, the threshold value for a bound state to appear $U_{min}$ is positive ($\delta K_{min} >0 $). Numerical calculation using DMM yields $U_{min} \approx 0.35 $. Following the line of reasoning presented in Ref.~\cite{our3}, we introduce the characteristic length scale $\xi$ as follows:
\be
  \xi^{-1} = q^\prime_D \frac{|U - U_{min}|}{U},
\ee
where $q^\prime_D \sim 1/a_0$. Next, the localized state frequency reads
\be
  \omega_{loc} = \omega_0 + \alpha \xi^{-2}.
\ee

Importantly, at $U \approx U_{min}$ we have $\xi \gg a_0$, and the phenomenon of resonant scattering emerges. In this regime the bulk expression for damping
\be \label{Tdamping}
\Gamma_\m{q} \sim \omega_\m{q} c_{imp} F \frac{\xi}{a_0} \frac{q \xi}{1 + (q \xi)^2}
\ee
reveals the drastic enhancement of the long-wavelength phonon line broadening, $\Gamma_\m{q} \sim 1/q$, which reflects the inapplicability of the first-order T-matrix approximation within the proximity of the resonance. In nanoparticles the finite size quantization $q \rightarrow q_n$ makes Eq.~\eqref{Tdamping} qualitatively correct. Furthermore, the more involved self-consistent theory (see Ref.~\cite{utesov2021}) results in $\Gamma_\m{q} \sim c^{2/3}_{imp} + O(q^2)$. In any case the damping in the resonant regime becomes large and the particle size dependence for overlapped levels $\Gamma \sim 1/L$ is violated. For Gaussian disorder with many nearly-resonant scatterers we indeed observed such a behavior numerically (see Fig.~\ref{ShellRes} below).

The eigenfunction $Y_{loc}$ of the bound state of phonon and strong impurity is mainly concentrated in the closest proximity of the impurity (see Fig.~\ref{Waves}(a)). Its characteristic size $a$ is of the order of several $a_0$. This allows to estimate the Raman intensity of the localized state. If $Y_0$ is the characteristic value of $Y_{loc}$, its normalization yields
\be
  \int Y_{loc}^2 dV \sim a^3 Y^2_0 \Rightarrow Y_0 \sim a^{-3/2}.
\ee
So, its Raman intensity can be estimated as follows:
\be
I_{loc} = \left| \int Y_{loc} dV \right|^2 \sim a^6 Y^2_0 \sim a^3,
\ee
whereas for usual ``volume'' states have $I \sim L^3$ (see Appendix). Thus, we make a conclusion that the contribution to Raman intensity stemming from surface localized states is small as compared to the volume modes contribution. When the amount of localized states is large their frequencies are distributed over a broad region, providing a ``pedestal'' for the broadened volume modes in the Raman spectrum (see below).

Another effect of bound states with strongly localized wave functions is a direct consequence of eigenstates orthogonality condition: these states ``repel'' other states from the corresponding regions of space. It . Indeed, for the smooth $n$-th eigenfunction (volume mode) we get
\be
  \int Y_n(\m{r}) Y_{loc}(\m{r}) dV \approx \int Y_n(\m{r}) \delta(\m{r} - \m{r}_i) dV = 0,
\ee
where $\m{r}_i$ is the impurity coordinate. The last equation implies $Y_n(\m{r}_i) = 0$ (which should be understood as the negligible amplitude of the volume mode eigenfunction in the vicinity of defect).

When the disorder strength in the nanoparticle shell is large enough, the localized states ``push'' the volume phonon modes away to the pure core, see Fig.~\ref{Waves}(b). Moreover, the strong impurities with negative $U$ also do not allow the volume modes to propagate, since their maximal optical phonon frequency is significantly smaller than $\omega_0$. Thus, the wave-functions of volume propagating modes (upon the averaging) acquire only exponentially decaying tails in the disordered shell. This is manifested in narrow pure core peaks in the particle Raman spectrum. Their positions are generally determined by the pure particle core size and shape (see Fig.~\ref{SurRam}(c)). If there is an additional disorder in the particle core it can be described using the approach of Refs.~\cite{our3,our4}.

Trying to apply the above qualitative picture for particles with small cores, we observe that the Raman intensity of volume modes becomes comparable with the one of localized states. In this case the entire particle should be considered as amorphous and our core-shell separation breaks down. The theory of RS in amorphous materials requires completely different approach. The corresponding RS cannot be treated as a set of weakly broadened pure particle phonon lines with their own Raman intensities, see Fig.~\ref{BulkAm}. The corresponding theory is out of the scope of the present paper.

\section{Results and Discussion}
\label{SRes}

In this section we compare our analytical quantitative and qualitative predictions with the results of numerical modeling, and discuss the observed physical picture. Both DMM or EKFG calculations have been performed for various problems at hands. In practice they give the same results on a qualitative level, and almost similar results on qualitative one. For instance, the threshold for bound state emergence is slightly different due to different ways of introducing the disorder in these two methods. The numerical approaches we use here for derivations of the line broadening parameters are discussed in details in Ref.~\cite{our4}.

\subsection{Bulk disorder}

When the bond disorder is introduced for the entire volume of the nanoparticle we can distinguish two well-understood regimes of broadening at small $S \ll 1 $, similarly to case of mass disorder~\cite{our3,our4}.

First, at smallest $S$, the regime of separated levels broadening appears. The disordered particle density of states has the structure of separated broadened lines, and their widths obeys the law:
\be
  \Gamma_n \sim \frac{\sqrt{S}}{L^{3/2}}.
\ee

When the disorder strength grows up, the spectral weights of individual phonon lines begin to overlap, which results in the bulk-like DOS. This situation can be discussed using the broadened phonon lines of the pure particle with the broadening given by
\be
  \Gamma_n \sim S q_n \sim \frac{S n}{L}.
\ee

The crossover scale between these two regimes can be introduced either in the form of crossover disorder strength $S_c$ or crossover particle size $L_c$
\be \label{cross}
  S_c \sim \frac{a_0}{L} \Longleftrightarrow L_c \sim \frac{a_0}{S}
\ee
with the material-dependent prefactor in the r.h.s. of Eq.~\eqref{cross}. For nanodiamonds, it is of the order of $0.1$ (see Ref.~\cite{our3} for details).

\begin{figure}[tbp]
\includegraphics[width=0.99\linewidth]{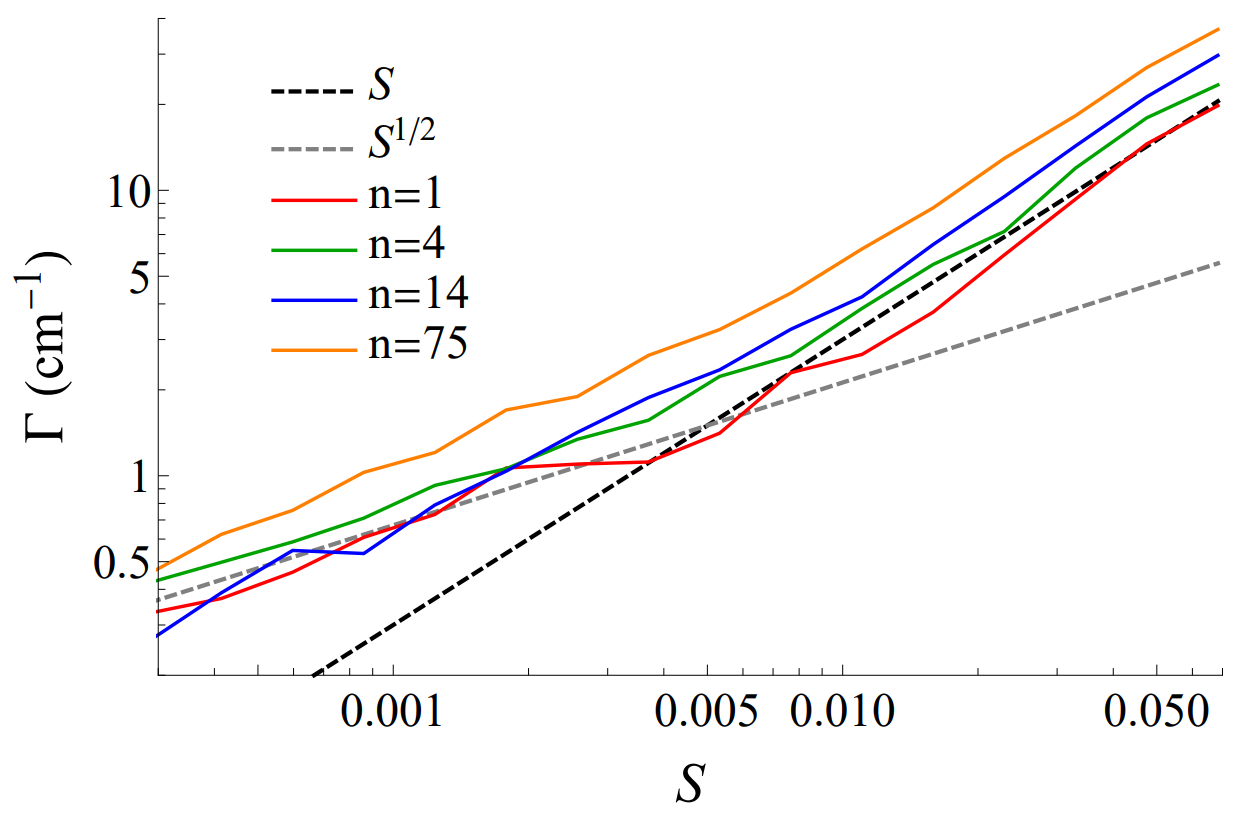}
\caption{The optical phonon linewidth $\Gamma_n$ versus $S$ for bond-disordered  2.4 nm spherical diamond  particles for modes with various quantum numbers obtained within DMM. The disorder is distributed over the entire particle volume. The asymptotes $\Gamma \sim S^{1/2}$ and $\Gamma \sim S$ for ``pure'' separated and overlapped regimes are also shown. }
\label{BulkS}
\end{figure}

Numerically we observe precisely this kind of behavior. Using DMM we study broadening of the optical phonon lines in $2.4$~nm diamond particles. The results are shown in Fig.~\ref{BulkS}. One can see (i) the crossover between the regimes of separated and overlapped levels in the region of $S \sim 0.001 \div 0.01$, (ii) relatively weak dependence of broadening on the phonon quantum number in the separated levels regime, and (iii) significant growth of the linewidth with quantum number when the levels are overlapped.

\begin{figure}[tbp]
\includegraphics[width=0.99\linewidth]{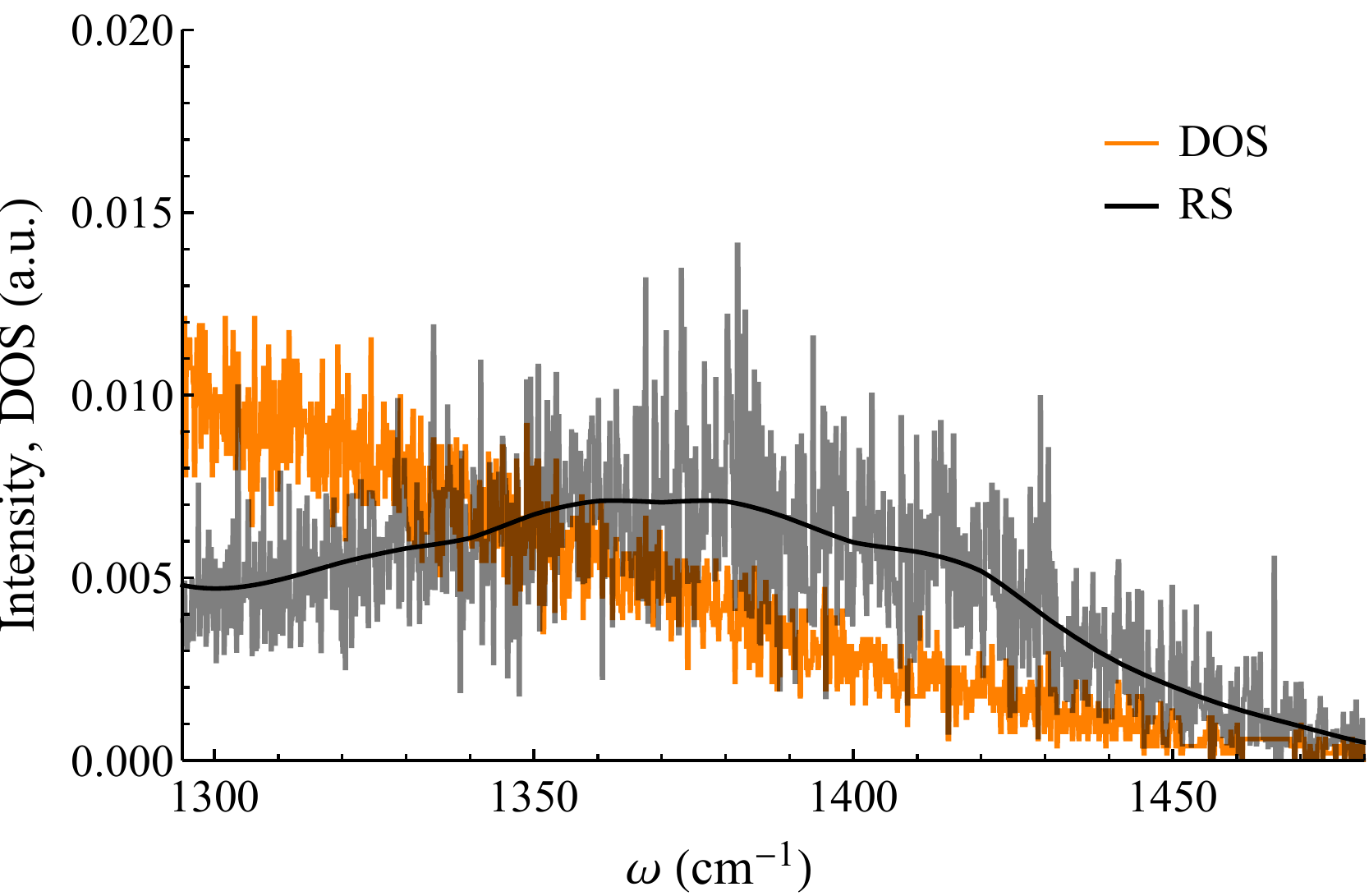}
\caption{The Raman spectrum (black line) and the density of states (orange line) obtained within the EKFG approach numerically for a strongly disordered (amorphous) cubic nanodiamond with the size $L=4.4$nm and with the disorder strength $S = 0.04$, the disorder is distributed over the entire particle volume. The description of the RS in terms of broadened (and shifted) individual phonon lines is incorrect. Cf. Fig. 4 from Ref.~\cite{our4}.}
\label{BulkAm}
\end{figure}

Finally, when the disorder becomes strong enough ($ \langle |\delta K| \rangle $ being of the order of the bound state threshold), the particle is essentially disordered and its description in terms of broadened pure particle phonon lines is incorrect, see Fig.~\ref{BulkAm}, where the results for the EKFG approach have been depicted (the DMM-BPM results are almost the same). One can see, that in this regime the individual phonon contributions to the RS are indistinguishable. \footnote{Similar RS smoothing with respect to the pure material takes place for bulk amorphous materials, e.g., for amorphous silicon \cite{smith1971raman,yogi2018quantifying} and for highly damaged by the ion implantation diamonds \cite{orwa2000raman}.} When there are many nearly resonant scatterers for pure particle modes, their effect cannot be described using the Born approximation, and the law $\Gamma \sim 1/L$ is violated (see Fig.~\ref{ShellRes} to observe similar effect illustrated within the core-shell model). Moreover, the T-matrix approach, which relies on small impurity concentration, is also inapplicable in this case. This strongly disordered regime (relevant to amorphous particles) and its quantitative theory in the confined geometry is a challenging problem, which lies beyond the scope of the present study. 

\subsection{Disordered shell and amorphization}

Next, we turn to the problem of bond disorder in the particle shell. Along with the conventional particle size $L$, there emerges a novel parameter $L_0$ -- the pure core size.

Here for disorder strength we use the definition~\eqref{dis2}. At $S_{\rm shell} \ll 1$ the difference between $L-L_0$ (disordered shell thickness) and $L$ leads to diminishing of the damping rates in comparison with the case of volume disorder ($L_0=0$). There are several important reasons for that: (i) the decrease of volume where scattering can take place, and (ii) the smallness of the eigenfunction $\m{Y}_n(\m{R}_l)$ amplitude near the boundary (which is evident, e.g., from the Dirichlet boundary conditions). The last statement is applicable at $L-L_0 \ll L$ for modes with $n \sim 1$. For phonons with large $n$ in nanoparticles the eigenfunctions are not too small even near the boundary, thus the corresponding linewidth $\Gamma_n$ grows with $n$ (see Appendix for more details). Importantly, this means that the first Raman-active mode can be separated from other modes, which belong to quasicontinuum and exist in the overlapped regime.

\begin{figure}[tbp]
  \includegraphics[width=0.99\linewidth]{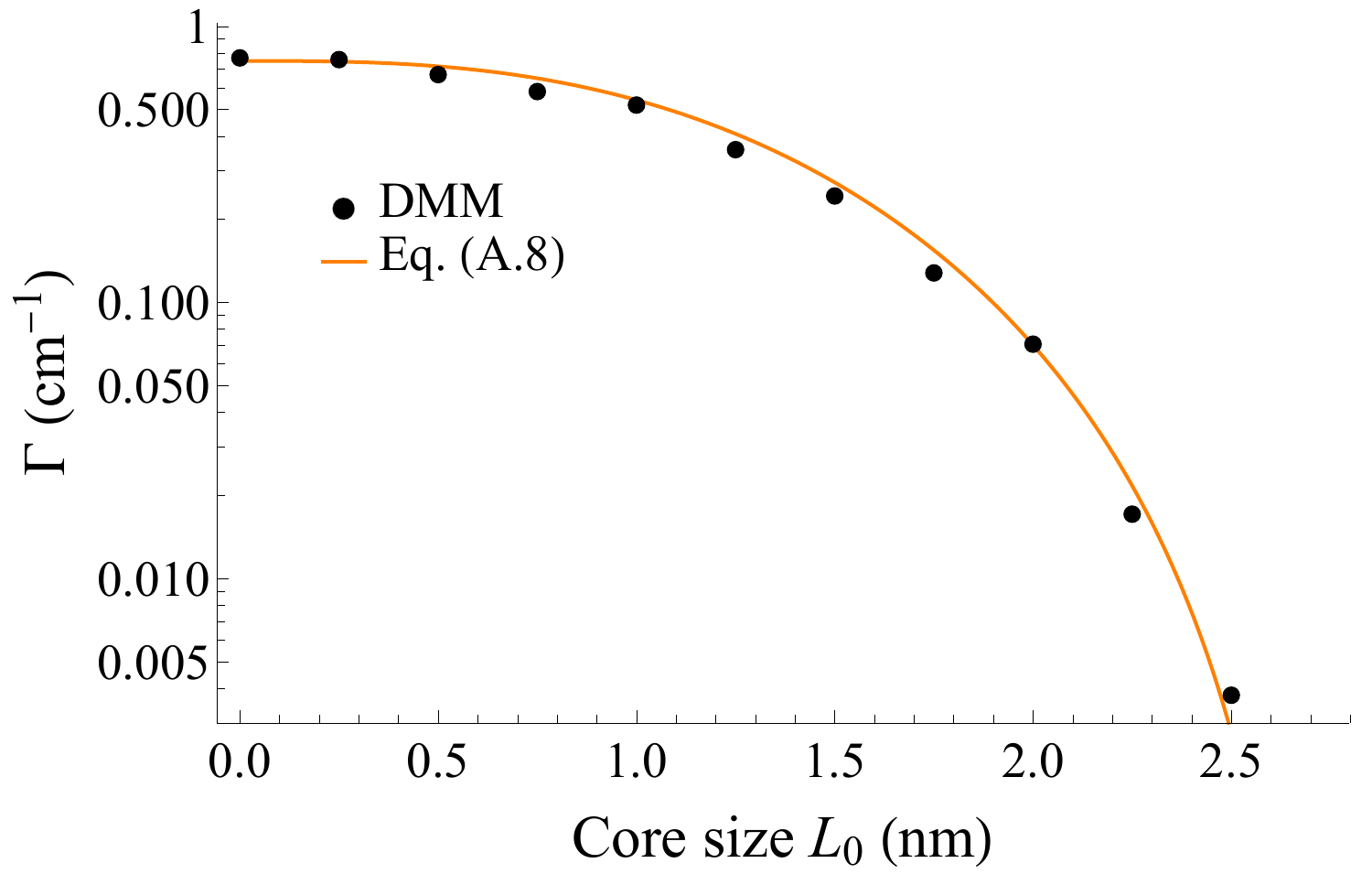}
  \caption{The damping of the first optical phonon mode $\Gamma_1$ for disordered shell with $S_{\rm shell} = 0.0005$ (separated levels), as a function of the disorder-free core size $L_0$. The total particle size is $L=2.75$nm. Black dots show the result of calculations within the DMM approach and the orange curve is for $f(L_0/L)$, given by Eq. \eqref{GL0}.}
  \label{figSmallS}
\end{figure}

\begin{figure}[tbp]
\includegraphics[width=0.99\linewidth]{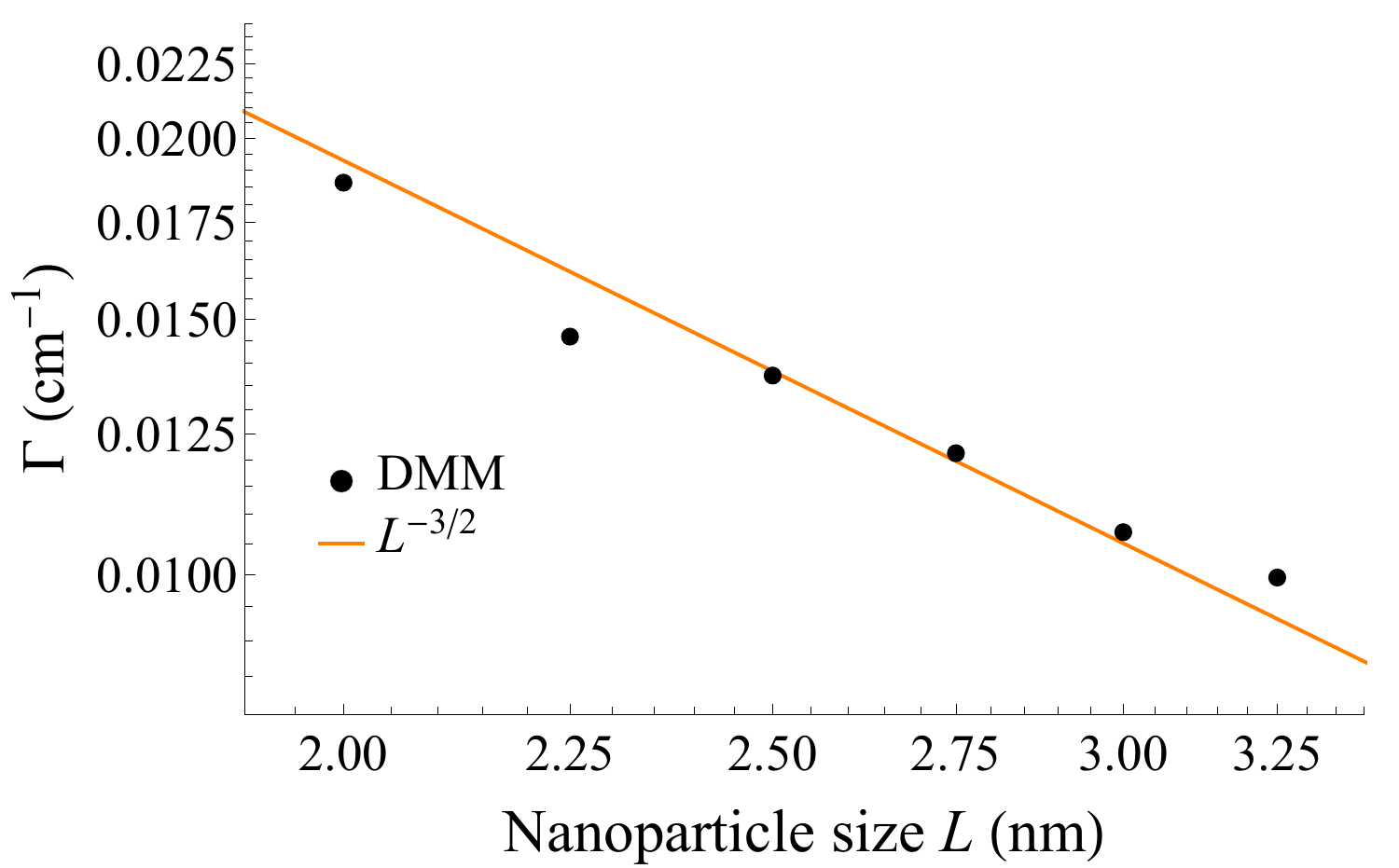}
\caption{The linewidth of the highest optical phonon mode $\Gamma_1$ as a function of the nanodiamond size calculated numerically for disordered shell with $S_{\rm shell}=0.002$. The ratio of the pure core size to the whole nanoparticle size is fixed, $L_0/L = 5/6$. The dots show the result of DMM calculations and the orange line is a guide for eyes depicting the power function with exponent -3/2, which corresponds to the regime of separated levels. The shapes of both particle and its core are the spheres.}
\label{figSmallS2}
\end{figure}

We illustrate this picture in Fig.~\ref{figSmallS}. Taking $S_{\rm shell} =0.0005$ (separated levels regime) we observe the substantial diminishing of the broadening parameter $\Gamma_1$ with the increasing of $L_0$ at fixed $L = 2.75$nm. Next, varying the particle size $L$ at fixed ratio $L_0/L$, we find the familiar $\Gamma \sim 1/L^{3/2}$ behavior (see Fig.~\ref{figSmallS2}). At last, fixing the shell width $L-L_0$, we see the faster diminishing of $\Gamma \sim 1/L^4$, which happens due to the decrease of eigenfunctions amplitudes in the disordered shell, see Eq.~\eqref{GL0} (cf. similar case of surface corrugations in Ref.~\cite{our4} Sec. VI).
 
\begin{figure}[tbp]
\includegraphics[width=0.99\linewidth]{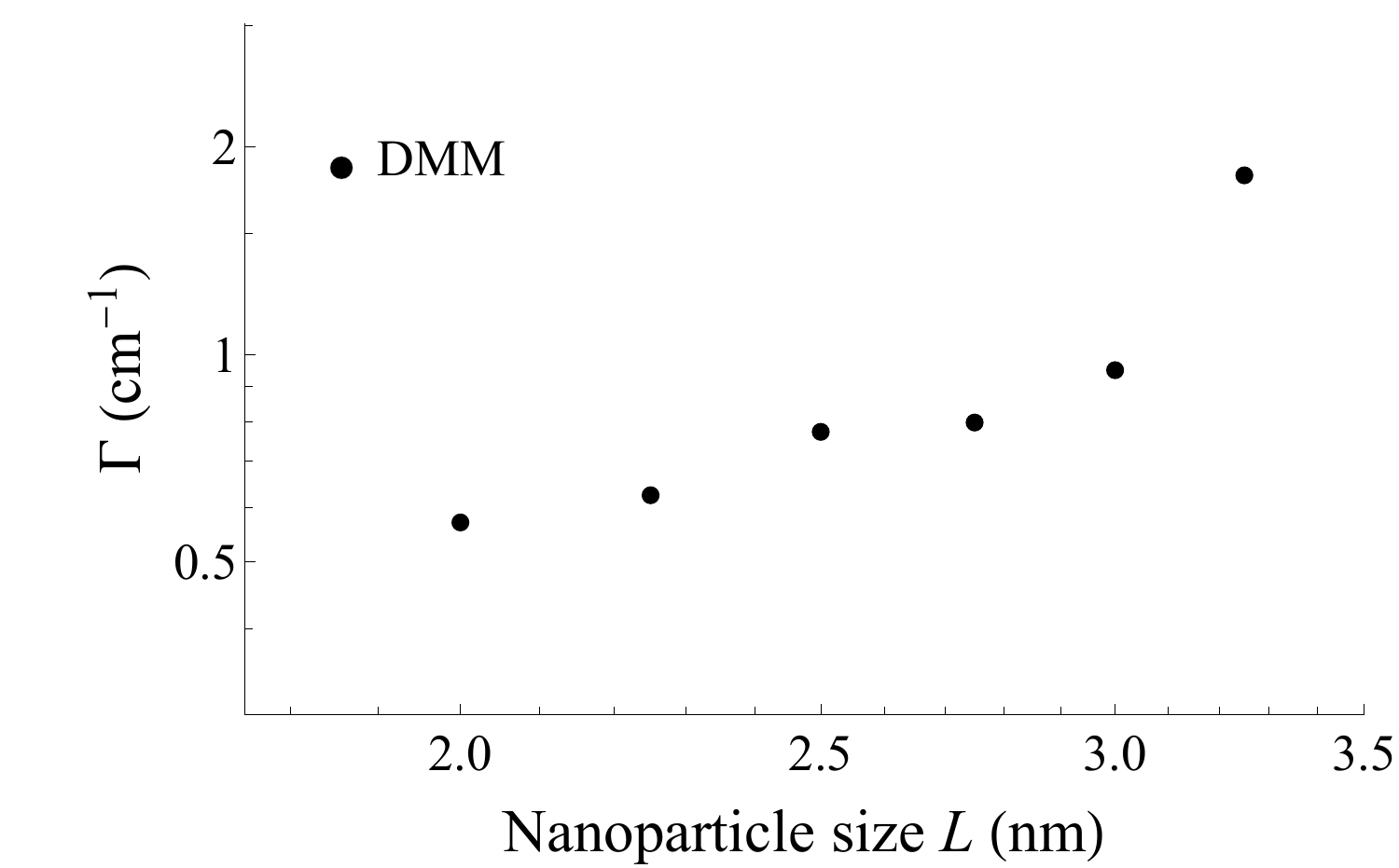}
\caption{Optical phonon linewidth of the first mode $\Gamma_1$ versus the particle size $L$ for the bond-disordered spherical diamond particles with disordered shell, pure core, and the core-to-size ratio $L_0/L = 5/6$ obtained within DMM.  Disorder strength $S_{\rm shell} = 0.047$, provides many resonant impurities in the shell. In particular, it leads to stronger broadening for larger particles. }
\label{ShellRes}
\end{figure}

When the disorder $S_{\rm shell}$ becomes strong enough, many bound states and resonant scatterers emerge. The volume modes broadening significantly increases. Importantly, this can happen even before small-$S$ overlapped levels regime appears (at least for narrow disordered shell). We investigate the corresponding effects numerically studying linewidth $\Gamma_1$ particle size dependence for parameters $S_{\rm shell}=0.047$ and $L_0= 5 L/6 $ (see Fig.~\ref{ShellRes}). Such a disorder strength is equivalent to $\langle |\delta K| \rangle \approx 0.22$, which provides a lot of nearly resonant impurities in the shell. As a result, we observe the violation of standard $1/L^{3/2}$ law for separated levels. Instead, one can see the growth of $ \Gamma_1$ when $L$ increases, as it is expected qualitatively for modes with finite quantized $q_n$ in the presence of resonant scatterers.

\begin{figure}[tbp]
\includegraphics[width=0.99\linewidth]{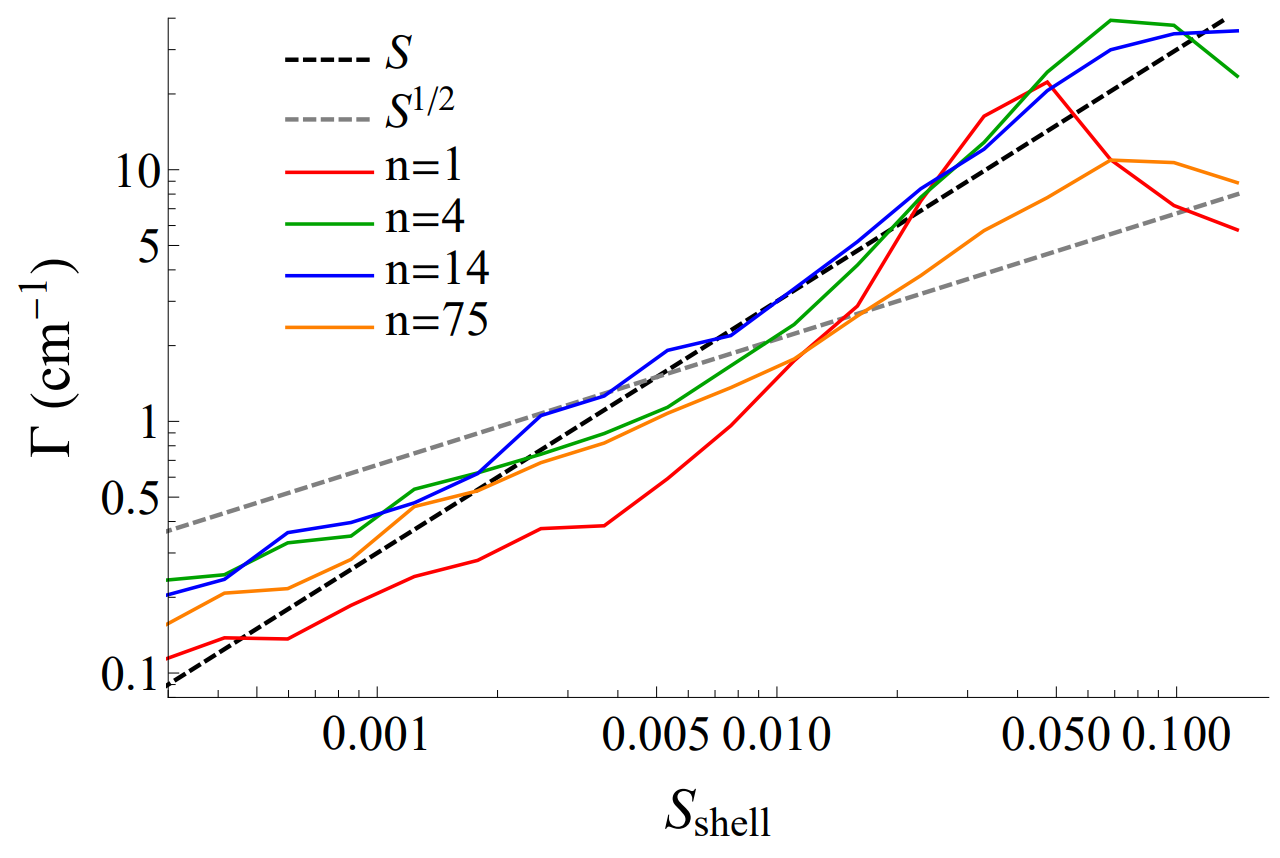}
\caption{The linewidths of several optical phonons  as functions of the disorder strength $S_{\rm shell}$ calculated for a $L=3$ nm spherical nanodiamond with the core-to-size ratio $L_0=L/2$ within the DMM approach.}
\label{figStrongS}
\end{figure}

\begin{figure}[th]
\includegraphics[width=0.9\linewidth]{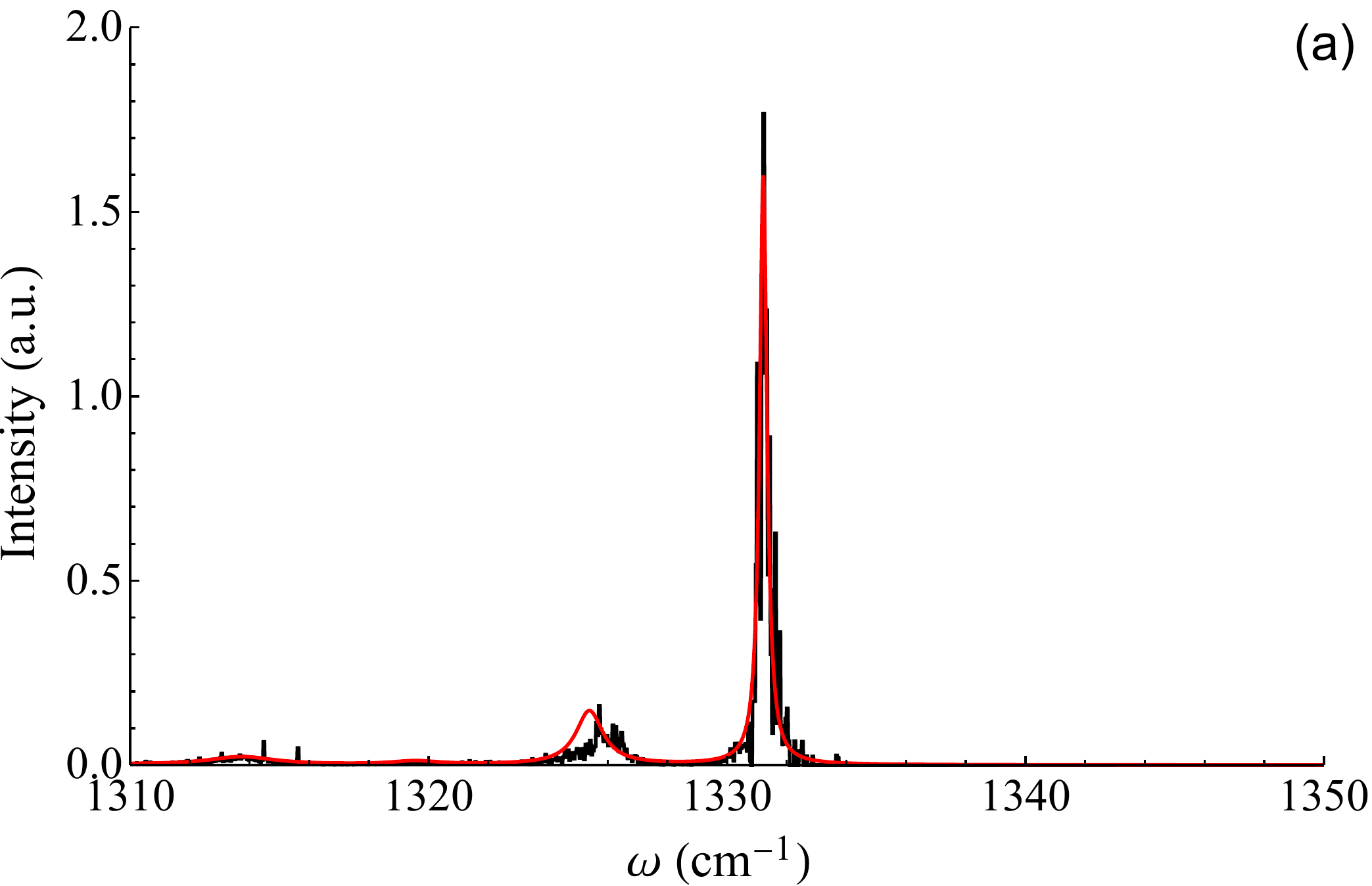}
\includegraphics[width=0.9\linewidth]{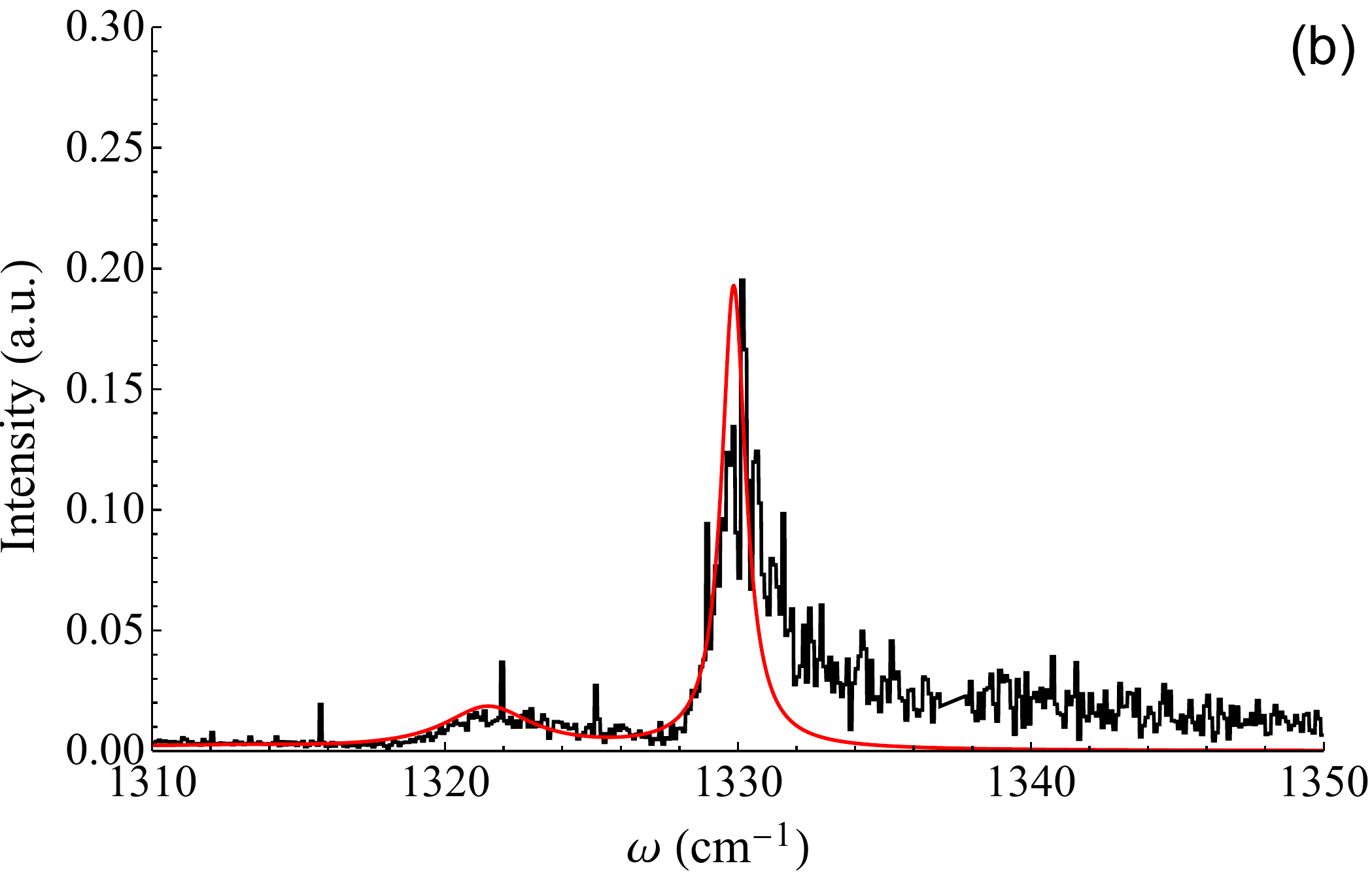}
\includegraphics[width=0.9\linewidth]{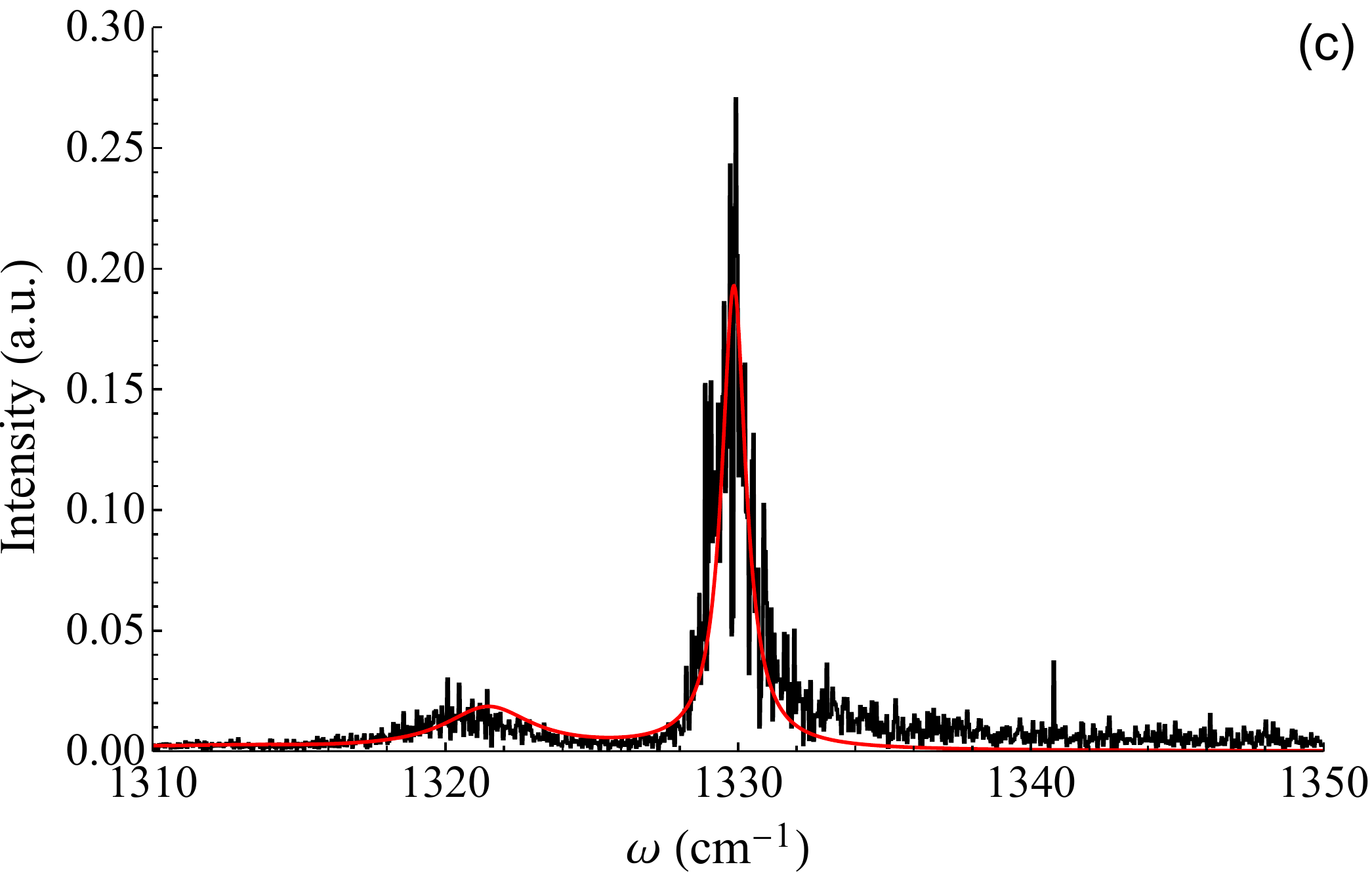}
\caption{The Raman spectra of cubic nanodiamonds with disordered shells ($S_{\rm shell}=0.0009, 0.01,0.09$ for (a), (b), and (c), respectively); the diamonds have the size $L \approx 5.3$ nm and the core-to-size ratio $L_0/L = 5/6$. The spectra are obtained numerically within the EKFG approach numerically (black lines). The main peak position in (a) is determined by the whole particle size $L$, whereas in (b) and (c) it corresponds to the size of the pure particle core $L_0$. The red line stands for the theoretical Raman spectrum as a sum of contributions from the broadened phonon lines of the particle with size $L$ (a) and $L_0$ (b,c). }
\label{SurRam}
\end{figure}

Under the further increase of disorder strengths ($S_{\rm shell} \sim 0.1$) we observe numerically the phenomenon of effective  ``localization'' of volume modes in the pure particle core discussed above. This statement is correct in average; usually there are fluctuations of disorder within the shell resulting in weakly disordered regions, where the volume modes can have significant amplitudes (see Fig.~\ref{Waves}(b)), which provides the effective broadening of these phonon lines.

In Fig.~\ref{figStrongS} we track the $S_{\rm shell}$-dependence of the linewidth for several optical phonon modes in $3$nm diamonds with the core size $L_0 = 1.5$nm. In the strongly disordered regime the broadening firstly drastically enhances upon increasing of $S$, then it saturates and starts to decrease.

Such a dependence has important consequences in the Raman spectra averaged over the particles ensemble. Using the EKFG approach we calculate these spectra for cubic diamonds of size $L \approx 5.3 \,$ nm with $L_0/L= 5/6$, see Fig.~\ref{SurRam}. At the smallest $S_{\rm shell}=0.0009$ (Fig.~\ref{SurRam}(a)) we observe the Raman spectrum, typical for weakly disordered particles of size $L$, where two peaks can be distinguished: the main peak from the highest $\m{n}= (1,1,1)$ phonon mode, plus the $11/3$ times broader (see Appendix) peak stemming from $\m{n}= (3,1,1)$ mode and equivalent modes.

When $S_{\rm shell} = 0.01$ (Fig.~\ref{SurRam}(b)) many localized in the shell states emerge. The peaks in RS, which correspond to broadened phonon lines of a particle with the size $L_0$ (quantized core modes), are well pronounced, however the shell signal is also signinficant. It has the shape of a long-lasting shoulder to the right from the main Raman peak.

Finally, at largest $S$ the shoulder stemming from the shell becomes somewhat negligible (see Fig.~\ref{SurRam}(c), where $S_{\rm shell}=0.09$). The RS can be described as the one of $L_0$-sized particles for which the shell plays the role of surface disorder. So, from the point of view of Raman spectroscopy such particles behave as the particles of size $L_0$, while the amorphous shell provides only a small correction to the whole particle RS, which stems from the localized bound states (see Fig.~\ref{Waves}) with small Raman intensities.

\section{Conclusions}
\label{SCon}

The present paper contributes to the theory of Raman spectra of disordered crystalline nanoparticles. We address the effect of disorder in interatomic bonds on the optical phonon lines broadening, which reveals the regimes of separated and overlapped levels. The former is characterized by the damping $\Gamma_n \sim \sqrt{S}/L^{3/2}$ weakly depending on the quantum number $n$ of the phonon mode, whereas in the latter $\Gamma_n \sim S/L$ and grows with $n$. When the disorder strength $S$ becomes large enough many localized states emerge and the particle becomes essentially disordered. The representation of the Raman spectrum as a superposition of contributions from broadened pure particle phonon lines fails in this case. This strongly disordered regime, which can be considered as a minimal model of amorphous particles and where the RS has a shape of a very broad peak, is out of the scope of the present paper.

We model the important in practice case of particles with amorphous surface within the framework of the core-shell model. Corresponding analysis reveals, along with the trivial separated levels regime of broadening at small $S_{\rm shell}$, an interesting behaviour of RS in the case of large $S_{\rm shell}$. The strongly disordered shell repels the optical phonon modes of the particle pure core. As a result, their broadening is rather weak because the shell plays a role of surface disorder. Consequently, well-resolved peaks originated from the phonon modes of the pure core are visible in the RS, while the amorphous shell contribution is broad and has a very small amplitude. Importantly, this means that when treating the RS of nanoparticles with amorphous coating using the method developed in Ref.~\cite{ourshort} one obtains the parameters of particle \textit{cores} (its mean size, size distribution function, disorder strength, and shape) rather than that of particles as a whole.

\begin{acknowledgments}

This work is supported by the Russian Science Foundation (Grant No. 19-72-00031).

\end{acknowledgments}

\appendix*

\section{EKFG approach and nanopowder Raman spectrum}
\label{SAppend}

\subsection{Raman spectrum}

Here we provide some important details necessary for understanding of our approach to the construction of the Raman spectrum for ensemble of nanoparticles (nanopowder). For the sake of simplicity we shall use the EKFG approach equations for cubic particles. General ideas are described in details in Refs.~\cite{ourDMM,ourEKFG,our3,our4}. 

EKFG eigenfunctions (see Eqs.~\eqref{EKFG1},~\eqref{EKFG2}, and~\eqref{EKFG3}) for cubic nanoparticle with the edge $b$  are given by the following expression:
\begin{equation}
  \label{WFcub}
  Y_{\bf n}= \left( \frac{2}{b} \right)^{3/2} \!\!\! e^{-i \omega t} \sin{\frac{\pi n_1 x}{b}} \sin{\frac{\pi n_2 y}{b}} \sin{\frac{\pi n_3 z}{b}},
\end{equation}
where the vector ${\bf n}=(n_1,n_2,n_3)$ enumerates the eigenstates, and the eigenvalues are
\begin{equation}
  \omega_\mathbf{n}=\omega_0 - \alpha  \left(\frac{\pi}{b} \right)^2 (n^2_1+n^2_2+n^2_3).
\end{equation}
The Raman intensity of each mode reads 
\be 
  I_\m{n} &=& \left| \int Y_\m{n} dV \right|^2 \\ &=& \frac{8^3 V}{\pi^6} \left( \frac{(n_1  \, \text{mod} \, 2) (n_2  \, \text{mod} \, 2)(n_3  \, \text{mod} \,2)}{n_1 n_2 n_3}\right)^2, \nonumber
\ee
where the symbol $(n_i \, \text{mod} 2)$ is the remainder of the division of $n_i$ by 2 (so, there are Raman-active and Raman-silent modes). Importantly, for this ``volume'' modes the Raman intensity is proportional to the particle volume $V$.

Next, assume that the source of disorder is known and one has an expression for the linewidth as a function of the phonon quantum number $n$ and the particle size $L$. So, Then the Raman spectrum of such particles is given by 
\be \label{RamanL1}
I_L(\omega) \propto \sum_n I_n \frac{\Gamma_n(L)}{(\omega - \omega_n)^2 + \Gamma^2_n(L)}.
\ee
For powder with the distribution function $f(L)$ the Raman spectrum reads
\be
  I(\omega) = \sum_L I_L(\omega) f(L).
\ee

\subsection{Disorder and phonon lines broadening}

Both the DMM and the EKFG approaches lead to the same diagram technique for the phonon linewidth calculation. The basic elements are the phonon Green's functions~\eqref{DQ} and the scattering vertex~\eqref{himp}. For weak Gaussian disorder phonon self-energy $\Pi_n(\omega)$ can be calculated using Born approximation (the self-consistent one in the case of separated levels, which is realized for weakest disorder and smallest particles), see Fig~\ref{Diagrams}.

The optical phonon eigenfunctions are small near the particle surface $\sim \pi n \delta l/L $, where $\delta l$ is the distance to the boundary (cf. Eq.~\eqref{WFcub}). So, the contribution to the self-energy from defects near the particle surface is much smaller, than the one from defects near the particle center:
\begin{equation}\label{SigmaW1}
  \Pi_n(\omega) = \frac{S_{\rm shell} \, V_0 \, \omega^3_0}{8} \int_{dis} d^3\mathbf{r} \sum_{n^\prime} \frac{Y^2_{n}(\mathbf{r}) Y^2_{n^\prime}(\mathbf{r}) }{\omega^2 - \omega^2_{n^\prime} - 2 \omega_{n^\prime} \Pi_{n^\prime}(\omega) },
\end{equation}
where the integration is restricted to the disordered region of particle. This is the reason why for particles with a thin disordered shell the effect of phonon linewidth broadening is negligibly small in comparison with the similar disorder distributed over the particle volume (see, e.g., Fig.~\ref{figSmallS}). For separated levels only the term with $n = n^\prime$ is important in sum~\eqref{SigmaW1}. Hence,
\be
  \Gamma_n \sim \sqrt{S_{\rm shell}} \left[ \int_{dis} d^3\mathbf{r} Y^4_{n}(\mathbf{r}) \right]^{1/2}.
\ee
Then, the diminishing of the damping with the growth of the core can be quantified (for the first phonon line of spherical particle) as follows:
\be \label{GL0}
\frac{\Gamma_1(L_0)}{\Gamma_1(0)} \equiv f(L_0/L) = \sqrt{1 - \frac{\int_{r \leq L_0/2} d^3\mathbf{r} Y^4_{1}(\mathbf{r})}{\int d^3\mathbf{r} Y^4_{1}(\mathbf{r})}}.
\ee

For phonon lines with higher quantum numbers the broadening is enhanced due to larger amplitude of eigenfunctions in the shell region (see Eq.~\eqref{WFcub}). For example, in the separated levels regime for cubic particle, Eq.~\eqref{SigmaW1} yields
\be
  \frac{\Gamma_{\m{n}}}{\Gamma_{\m{1}}} = \frac{n^2_1 + n^2_2 + n^2_3}{3}.
\ee
In spherical particles its counterpart for the Raman-active modes simply yields $\Gamma_n/\Gamma_1 = n^2$, where $n$ is the principal quantum number.

When the disorder strength $S$ is not small the Born approximation fails. Moreover, due to effectively large concentration of impurities the T-matrix approach, suitable for rare strong impurities, also cannot be applied. The nanoparticles are essentially disordered, and the description of Raman spectra in terms of a superposition of broadened peaks stemming from the pure particle phonon modes (see Eq.~\eqref{RamanL1}) also becomes incorrect. There are many localized (on strong bonds with $\delta K / K > 0.35$) states that governs the RS in this regime.

Nevertheless, we analyze corresponding Raman spectra numerically. For each disorder realization one has a set of eigenfunctions $Y_n$, eigenfrequencies $\omega_n$, and corresponding Raman intensities $I_n$. Next, this data should be averaged over disorder realizations. In practice, we average over 128 samples in our calculations.

\bibliography{KFG}

\end{document}